\documentstyle[11pt,moriond,psfig]{article}

\def\Quadrat#1#2{{\vcenter{\hrule height #2
  \hbox{\vrule width #2 height #1 \kern#1
    \vrule width #2}
  \hrule height #2}}}
\def\dAl{\mathop{\kern 1pt\hbox{$\Quadrat{8pt}{0.4pt}$} \kern1pt}}
\def\journal#1#2#3#4{{#1} {\bf #2}, #3 (#4)}
\def\prl{\em Phys. Rev. Lett.}
\def\prd{{\em Phys. Rev.} D}
\def\mnras{\em Mon. Not. R. Astron. Soc.}
\def\apj{{\em Astrophys. J.}}

\def\vp{\varphi}

\def\be{\begin{equation}}
\def\ee{\end{equation}}
\def\bea{\begin{eqnarray}}
\def\eea{\end{eqnarray}}
\begin{document}
\vspace*{4cm}
\title{Timing Effects of Gravitational Waves from Localized Sources}

\author{Sergei M. Kopeikin \footnote{
On leave from: {\em ASC FIAN, Leninskii Prospect, 53, Moscow 117924,
Russia}}}
\address{Friedrich-Schiller-Universit\"at, TPI, Max-Wien-Platz 1, D - 07743, Jena, Germany\\
Max-Planck-Institut f\"ur Radioastronomie, Auf dem
H\"ugel 69, 53121 Bonn, Germany}
\maketitle
\abstracts{Localized astronomical sources like a double
stellar system, rotating neutron star, or a massive black hole at the center 
of the Milky Way emit periodic gravitational waves. For a long time only a 
far-zone contribution of gravitational fields of the localized sources 
(plane-wave-front approximation) were a matter of theoretical analysis. 
We demonstrate how this analysis can be extended to take into account 
near-zone and intermediate-zone contributions as well. The formalism is used to
calculate gravitational-wave corrections to the Shapiro time delay in binary
pulsars and low-frequency (LF) pulsar timing noise produced by an ensemble of double
stars in our galaxy.}
\section{Introduction}
The extremely high precision of current pulsar timing 
observations
demands a better theoretical treatment of secondary effects in the
propagation of electromagnetic signals in variable gravitational fields
of oscillating and precessing stars, or binary systems. Especially important 
is the problem of propagation
of light rays in the field of gravitational waves emitted by a localized source
of gravitational radiation. 
A consistent approach for a complete and exhaustive solution of this problem 
has been developed in papers \cite{1}$^{-}$\cite{3} in the first 
post-Minkowskian 
approximation of General Relativity. 
We have demonstrated for 
the first time that the equations of light propagation in the retarded 
gravitational field of an arbitrary localized source emitting  
gravitational waves can be integrated exactly in closed form.  
It allows to examine the influence 
of the time-dependent gravitational field on the light propagation 
not only in the wave zone but also in cases when light passes through 
the intermediate and near zones of the source.
We have obtained explicit analytic expressions for light
deflection and integrated time delay (Shapiro effect) 
accounting for all possible retardation effects and arbitrary
relative locations of the source of gravitational waves, the 
source of light rays , and the observer. 
Coordinate dependent terms in the
expressions for observable quantities were singled out and used for 
the unique interpretation of observable effects. 
Our exploration essentially extends previous results regarding propagation of
light rays in the field of a 
plane monochromatic gravitational wave (see, e.g., papers \cite{4}$^{-}$\cite{7})
and significantly
surpasses theoretical approaches of other authors \cite{8}$^-$\cite{12}.
In the present paper we briefly discuss the  
developed formalism and apply it to
describe the Shapiro time delay in binary pulsars as well as to
estimate the intensity of LF timing noise produced by an ensemble of double stars of our galaxy.

\section{Propagation of light in time-dependent gravitational fields }

We solve the Einstein equations in the first post-Minkowskian approximation where
the metric tensor $g_{\alpha\beta}$ is decomposed linearly into the Minkowski metric 
$\eta_{\alpha\beta}$ and a small perturbation $h_{\alpha\beta}$,
\be
g_{\alpha\beta}= \eta_{\alpha\beta}+h_{\alpha\beta}\;.
\label{1}
\ee
The Einstein equations for $h_{\alpha\beta}$ in harmonic gauge read \footnote{In 
what follows we use geometrical units in which $G=c=1$.}
\be
\label{2}
\dAl h_{\alpha\beta}(t, {\bf x})=-16\pi S_{\alpha\beta}(t,{\bf x})\;,
\quad\quad
\mbox{where}\quad\quad
S_{\alpha\beta}(t, {\bf x})=T_{\alpha\beta}(t, {\bf x})-
\frac{1}{2}\eta_{\alpha\beta}\;T_{\;\;\lambda}^{\lambda}(t, {\bf x})\;. 
\ee
Herein $T^{\alpha\beta}$ is the tensor of energy-momentum of a system of 
massive particles \cite{13}
\be
\label{4}
T^{\alpha\beta}(t, {\bf x})=\sum_{a=1}^N 
\hat{T}_{a}^{\alpha\beta}(t)\delta\left({\bf x}-{\bf
x}_a(t)\right)\;,\quad\quad\quad\quad\quad
\hat{T}_{a}^{\alpha\beta}(t)=m_a \gamma_a^{-1}(t)u_a^\alpha(t)
u_a^\beta(t)\;,
\ee 
where $t$ is coordinate time, ${\bf x}=x^i=(x^1,x^2,x^3)$ denotes spatial
coordinates of a current point in space, $m_a$ and ${\bf x}_a(t)$ are the  
rest mass and spatial coordinates of the $a$-th particle, ${\bf v}_a(t)=
d{\bf x}_a(t)/dt$, $\gamma_a(t)=[1-v_a^2(t)]^{-1/2}$ is the  
Lorentz factor, $u_a^\alpha(t)=\{\gamma_a(t),\;\gamma_a(t){\bf
v}_a(t)\}$ is the
four-velocity of the $a$-th particle, $\delta({\bf x})$ is the  
3-dimensional Dirac
delta-function.
The solution of (\ref{2}) is the retarded {\it Li\'enard-Wiechert} tensor 
potential
\begin{eqnarray}
\label{5}
h^{\alpha\beta}(t, {\bf x})&=&4\sum_{a=1}^N\;\frac{\hat{T}_{a}^{\alpha\beta}(s)-
\frac{1}{2}\eta^{\alpha\beta}\hat{T}_{a\lambda}^{\lambda}(s)}
{r_a(s)-{\bf v}_a(s)\cdot {\bf r}_a(s)}\;,
\end{eqnarray}
where the retarded time $s=s(t,{\bf x})$ for the $a$-th body 
is a solution of the light-cone equation $s+|{\bf x}-{\bf x}_a(s)|=t$.
Here it is assumed that the field is measured at time $t$ and at the 
point ${\bf x}$. 

\begin{figure*}
\centerline{\psfig{figure=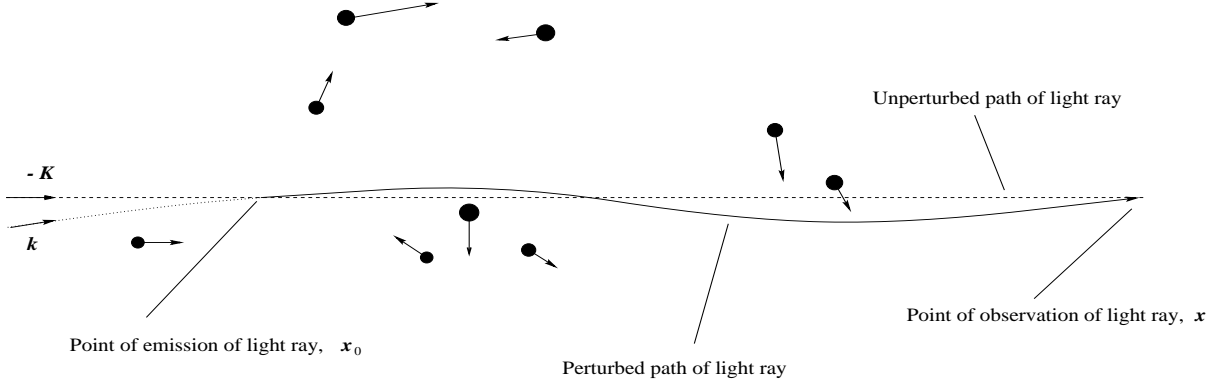,angle=270,height=5 cm,width=16 cm}}
\caption{Illustration of the light-ray's propagation history.
The light ray is emitted at the instant of time $t_0$ at the point ${\bf
x}_0$ and arrives at the point of observation ${\bf x}$ at the instant of time $t$.
Light-deflecting bodies moves along accelerated world lines during the time of
propagation of the light ray; their velocities at some intermediate instant
of time are shown by black arrows. In the absence of the light-ray-deflecting 
bodies the light ray would propagate along an unperturbed path (dashed line) which
is a straight line passing through the points of emission, ${\bf x}_0$, and
observation, ${\bf x}$. Direction of the unperturbed path is determined by the
unit vector ${\bf K}=-({\bf x}-{\bf x}_0)/|{\bf x}-{\bf x}_0|$. 
In the presence of the light-ray-deflecting bodies the light ray
propagates along the perturbed path (solid line). The perturbed trajectory of
the light ray is bent and twisted due to the gravito-electric (mass-induced)
and gravito-magnetic (velocity-induced) fields of the bodies. The initial boundary
condition for the equation of light propagation is determined by the unit
vector ${\bf k}$ defined at past null infinity by means of a dynamical backward-in-time
prolongation (dotted line) of the perturbed trajectory of light from 
the point of emission ${\bf x}_0$ in such a way that the tangent vector 
of the prolongated trajectory
coincides with that of the perturbed light-ray's trajectory at the point
of emission. The relationship between unit vectors ${\bf k}$ and ${\bf K}$
includes relativistic bending of light. }     
\label{covariant1}
\end{figure*}

We consider the motion of a light ray (photon) in the background 
gravitational field described by the metric (\ref{5}). The motion of the photon 
is defined by solving equations of a light geodesic (see Fig \ref{covariant1}).
The original equations of propagation of light rays are rather 
complicated \cite{2}. They can be simplified and
reduced to a form which allows to resort to 
a special approximation method for their integration. 
The reduced equations and details of the integration 
procedure can be found in our paper \cite{3}. In what
follows we discuss some of physical applications of the mathematical technique.

\section{Shapiro time delay in binary pulsars}

Damour \& Esposito-Far\`{e}se \cite{12} argued about the contribution of
gravitational waves emitted by a binary pulsar to the observed Shapiro time 
delay. As a consequence of the main result of their work they claimed that the 
contribution probably should be small. However, they were not able to
present conclusive mathematical arguments which are given in this section.

The general formalism of our paper \cite{3} yields for the 
time of propagation of the light-ray
\begin{eqnarray}
\label{qer}
t-t_0&=&|{\bf x}-{\bf x}_0|+\Delta(t,t_0)\;,
\end{eqnarray}
where $|{\bf x}-{\bf x}_0|$ is the Euclidean distance between the points
of emission, ${\bf x}_0$, and observation, ${\bf x}$, of the photon, and 
$\Delta(t,t_0)$ is the Shapiro time delay due to the gravitational field
of moving bodies 
\begin{eqnarray}
\label{shapd}
\Delta(t,t_0)&=&\frac{1}{2}k_{\alpha}k_{\beta}B^{\alpha\beta}(\tau)-
\frac{1}{2}k_{\alpha}k_{\beta}B^{\alpha\beta}(\tau_0)\;=\;2\sum_{a=1}^2
m_a\;B_a(s,s_0)\;,
\end{eqnarray}
\bea
\label{integral1}
B_a(s,s_0)&=&
\displaystyle{\int^{s}_{s_{0}}}\frac{[1-{\bf k}\cdot{\bf
v}_a(\zeta)]^2}{\sqrt{1-v^2_a(\zeta)}}
\frac{d\zeta}{t^{\ast}+{\bf k}\cdot{\bf x}_a(\zeta)-\zeta}\;,
\eea 
where the retarded time $s$ is obtained by solving the equation of light cone for 
the time of observation, the time $s_0$ is found by solving the same 
equation written down for the time of emission of photon $
s_0+|{\bf x}_0-{\bf x}_a(s_0)|=t_0$, and $t^\ast$ is the (constant) time of the closest
approach of the light ray to the origin of the coordinate system used for
calculations. 
Choosing the origin of the coordinate system to be the barycenter of the binary
pulsar, expanding (\ref{integral1}) in powers of $v_a/c$, and performing
integration result in \cite{3}
\begin{eqnarray}
\label{ert}
\Delta(t,t_0)&=&-2m_c(1-{\bf k}\cdot{\bf v}_c)
\ln\left[\left(r+{\bf k}\cdot {\bf r}\right)-
{\bf v}_c\cdot{\bf r}+({\bf k}\cdot {\bf r})
({\bf k}\cdot{\bf v}_c)\right]\;,
\end{eqnarray}
where $m_c$ is the mass of the pulsar's companion, ${\bf r}$ is the 
radial distance between the pulsar and companion,
${\bf v}_c$ is the companion's barycentric velocity, and all quantities are taken
at the barycentric time of emission of the pulsar's radio pulse. 
Formula (\ref{ert})
apparently demonstrates that the
velocity-dependent terms present in the Shapiro effect are
small and only marginally detectable in timing of binary pulsars. 
This proves the intuitive guess of the
paper \cite{12}. Formula (\ref{ert}) can be implemented in the parameterized
post-Keplerian formalism \cite{16}$^,$\cite{17}. Its explicit 
formulation is given in \cite{3}. In case of circular orbits it reads
\bea
\label{circ}
\Delta_S &=& -\frac{2G m_c}{c^3}\ln\biggl\{1-\sin i\left[
\sin\left(\vp-\vp_0\right)+\frac{x}{P_b}\frac{m_p}{m_c}\;\pi
\sin 2\left(\vp-\vp_0\right)\right]\biggr\}\;,
\eea
where $m_p$ is the pulsar's mass, $\vp$ is the orbital phase, 
$\vp_0$ is constant, $x$ is the projected semi-major axis of the
orbit, $P_b$ is the orbital period, and we have restored $G$ and the speed of 
light $c$ for convenience. The correction to the Shapiro delay given in
(\ref{circ}) may be observed in binary pulsars with orbits visible nearly 
edgewise.

\section{Low-frequency pulsar timing noise}

Each double star in our galaxy emits rather weak LF gravitational 
waves. However, an
ensemble of double stars may produce a noticeable amount of the gravitational
radiation which could be detected through the analysis of the long-term behavior 
of the timing residuals (see Fig. 2). 
Time variations 
caused by quadrupolar gravitational waves from a double star read as \cite{2}
\bea
\label{res}
\Delta(t, t_0)&=&\delta(t,r,\theta)-\delta(t_0, r_0,\theta_0)\;,
\eea
\be
\label{rr}
\delta(t,\theta,r)=\frac{k^i k^j}{1-\cos\theta}\left[
\frac{\dot{\cal I}_{ij}^{TT}(t-r)}{r}+
\frac{1}{1-\cos\theta}\;\frac{{\cal
I}_{ij}^{TT}(t-r)}{r^2}+\frac{A(\theta,t-r)}{r^2}\right]+O\left(r^{-3}\right)\;,
\ee 
where $k^i$ is the unit vector from the pulsar towards the observer,
$r$ is the distance `observer-double star'; $r_0$ is the distance 
`pulsar-double star' ($r_0=(r^2+R^2-2r R\cos\theta)^{1/2}$); 
$\theta$ is the angle between the directions `observer-double star' and
`observer-pulsar'; $\theta_0$ is the angle between the directions `pulsar-double
star' 
and `observer-pulsar'; ${\cal I}_{ij}^{TT}$ is the transverse-traceless quadrupole
moment of the double star; $A(\theta,t-r)$ is a regular function. If one takes 
a particular event of a circular orbit the formula
(\ref{rr}) is brought to the form
\bea
\label{ii}
\delta(t,\theta,r)&=&\frac{(G{\cal M})^{5/3}}{c^4\;\omega^{1/3}}
\frac{\cos^2\displaystyle{\frac{\theta}{2}}}{r}\left[
(1+\cos^2 i)\cos 2\Omega \sin 2\vp+2 \cos i \sin 2\Omega \cos 2\vp\right]\\
\nonumber\\\nonumber&&\hspace{-1cm} 
-\frac{(G{\cal M})^{5/3}}{4c^3\;\omega^{4/3}}\frac{\cot^2\displaystyle{
\frac{\theta}{2}}}{r^2}\left[
(1+\cos^2 i)\cos 2\Omega \cos 2\vp-2 \cos i \sin 2\Omega \sin 2\vp\right]+
O\left(\frac{G^{5/3}{\cal M}^{5/3}}{c^2\omega^{7/3}r^{3}}\right)\;,
\eea
where $\vp=\omega t+\sigma$ is the orbital phase of the double star, 
$\sigma$ is constant, $\omega=2\pi\nu$ is the orbital frequency, $\Omega$ is the
longitude of the ascending node, and ${\cal M}=(m_1 m_2)^{3/5}(m_1+m_2)^{-1/5}$ is
the "chirp" mass of the double star. 

An ensemble of double stars generates the gravitational wave noise with the 
autocovariance function ${\cal R}(\tau)$  
defined as the ensemble average 
of products of the time delay (\ref{res})  
\bea
\label{oo}
{\cal R}(\tau)&=&<\Delta(t,t-R)\;\Delta(t+\tau,t-R+\tau)>\;,
\eea
where we have used an approximation $t_0=t-R$. The ensemble averaging is defined
with respect to the set of random variables $\Omega$, $i$, $\vp_0$,
$\nu$, $\theta$, and $r$ \footnote{We note that variables $r_0$ and $\theta_0$ are not
independent and can be expressed through $r$ and $\theta$}. We assume that
the variables $\Omega$, $i$, $\vp_0$ are distributed uniformly. The probability 
distribution function of other variables of the ensemble is determined by 
main-sequence galactic 
double stars which occupy the frequency band \cite{19} from $10^{-13}$
to $10^{-5}$ Hz with the frequency and spatial distributions $F(\nu)$ and 
$n({\bf r})$ respectively \footnote{Function $F(\nu)$ is nomalized to unity, and
$n({\bf r})=n(\theta,r)$ is normalized to the total number of double stars  
in the ensemble. }
\be
\frac{d N(\nu,{\bf r})}{d\nu\;d{\bf r}}=F(\nu) n({\bf
r})\;,\quad\quad\quad\quad F(\nu)=\frac{\kappa(\nu)}{\nu}\;,
\ee
where it has been assumed that $F(\nu)$ and $n({\bf r})$ are statistically
independent. Function $\kappa(\nu)$ is almost constant over a wide range of 
frequencies. Hence,
we take $F(\nu)\simeq 0.054/\nu$. 
\begin{figure*}
\label{smk}
\centerline{\psfig{figure=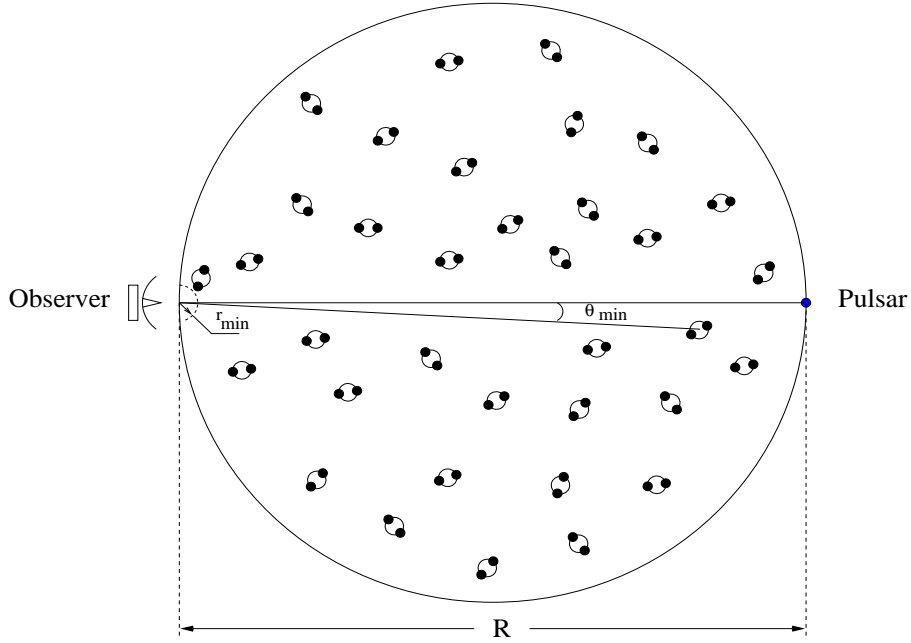,angle=270,height=8.5 cm,width=12 cm}}
\caption{A model of an ensemble of double systems producing LF gravitational
noise. The minimal distance observer-double star is denoted as $r_{min}$ and the
minimal angle between directions to the pulsar and a double star is denoted as
$\theta_{min}$. The distance from the observer to the pulsar is $R$. Concentration, $n$, of
double stars is assumed to be constant. Then, 
$r_{min}\sim n^{-1/3}$, $\theta_{min}\sim r_{min}/R$, and $n({\bf r})=3n/(4\pi
R^3)$.}
\end{figure*}
One can prove that the double-star-gravitational-wave noise is stationary with
spectrum $S(\nu)$ defined by the Fourier transform of the autocovariance function.
The spectrum has the following specific properties
\begin{itemize}
\item the noise consists of seven components having spectra
$S(\nu)\sim\nu^{-5/3-\alpha}$ ($\alpha=0\div 6$);
\item the amplitude of any component of the noise crucially depends on the spatial
distribution of galactic double stars between the observer and the pulsar as well
as in the neighborhood of each (e.g. a pulsar in a globular cluster);
\item the component of $S(\nu)$ with $\alpha=0$ is produced by the far zone
gravitational field ($\sim r^{-1}$) of double stars (pure
gravitational waves) and is regular everywhere including the line of sight of 
observer towards the pulsar, when $\theta=0$;
\item the component of $S(\nu)$ with $\alpha=2$ is due to 
the semi-far zone gravitational field ($\sim r^{-2}$) of double stars
and diverges along the line of
sight of observer towards the pulsar \footnote{This divergency of the spectrum
is closely related to the effect of augmentation of the time delay by 
gravitational lensing by the intervening matter \cite{2}};
\item the components of $S(\nu)$ with $\alpha=4,6$ are due to the intermediate zone ($\sim
r^{-3}$) and near zone ($\sim r^{-4}$) gravitational fields of double stars
and are regular everywhere.  
\end{itemize}
In order to give an estimate of the expected amount of the gravitational wave noise in
pulsar timing we use a simple model of uniform distribution of double stars in
the spherical domain
between the pulsar and observer at the Earth (see Fig. 2 for more
details). The strength of the timing noise is most conveniently characterized by
the, so-called, $\sigma_z$ statistic \cite{20} defined as
\bea\label{tay}  
\sigma_z(\tau)&=&\frac{\tau^2}{2\sqrt{5}}<c_3^2(\tau)>^{1/2}\;,
\eea
where $c_3(\tau)$ is a random variable being
proportional to the statistical fluctuations of 
the third time derivative of the pulsar's
timing residuals, $r(t)$, and $\tau$ is the total span of observational time. 
In the case of a single double star, 
${\em r}(t)=\Delta(t,t-R)$. 
In what follows we simplify the problem by 
accounting for only the line-of-sight divergent component of $\Delta(t,t-R)$ 
given in the second line of (\ref{ii}). Performing integration over the
ensemble's variables and
over the sperical volume \footnote{Origin of the spherical coordinate system is at
observer. Limits of integration w.r.t. to the radial
coordinate, r, and the declination, $\theta$, are ($r_{min}, R\cos\theta$), and
($\theta_{min},\pi/2-\theta_{min}$) respectively.} shown in Fig. 2 gives 
\be
\label{mm}
<c_3^2(\tau)>\simeq(2\pi)^6\int_0^{1/\tau}\nu^6 S(\nu)
d\nu\;,\quad\quad\mbox{with}\quad\quad S(\nu)\sim \nu^{-11/3}\;,\quad\quad\mbox{so
that}
\ee
\bea
\label{sig}
\sigma_z(\tau)&\simeq&0.7\cdot 10^{-18}\left(\frac{\cal M}{{\cal
M}_\odot}\right)^{5/3}\left(\frac{n}{{\rm 1}\;{\rm pc}^3}\right)
\left(\frac{R}{\mbox{1 kpc}}\right)
\left(\frac{\tau}{\mbox{1 yr}}\right)^{1/3}\;.
\eea
This limit is lower than that presently accessible in the timing of millisecond 
pulsars. However, we emphasize that the estimate (\ref{sig}): 1) is  model 
dependent, and 2) contribution of several other components of $\Delta(t,t-R)$ were
neglected. It is highly desirable to repeat our calculations with more 
realistic ensemble of
double star distributions and the complete account for all constituents of 
$\Delta(t,t-R)$. We argue that proceeding in this way the origin of the
incomprehensible `red' timing noise
discovered in PSR B1937+21 \cite{21} may be explained as a result of the 
LF gravitational
wave noise from double stars in our galaxy - a challenge both for theorists and
observers. 
\section*{Acknowledgments}
I am grateful to G. Neugebauer (FSU, Jena) and R. Wielebinski (MPIfR, Bonn) 
for hospitality and support of this work. I thank W.A. Sherwood for 
critical reading of the manuscript and 
valuable remarks and B. Paczy{\'n}ski for discussion.
 
\end{document}